\documentclass[prd, twocolumn, nofootinbib]{revtex4}
\usepackage{graphicx}
\usepackage{dcolumn}
\usepackage{epsfig} 
\usepackage{amsmath}
\usepackage{amsfonts}
\usepackage{graphicx}
\usepackage{amssymb}
\begin{document}
\title{Primordial Bispectrum and Trispectrum Contributions to the Non-Gaussian Excursion Set Halo Mass Function with Diffusive Drifting Barrier} 
 
\author{Ixandra E. Achitouv$^{\star,\dagger}$ }

\author{Pier Stefano Corasaniti$^{\star}$}

%\author{James G. Bartlett$^\dagger$}

\address{$^\star$ Laboratoire Univers et Th\'eories (LUTh), UMR 8102
 CNRS, Observatoire de Paris, Universit\'e Paris Diderot, \\ 5 Place Jules Janssen, 92190 Meudon, France}
 \address{$^\dagger$ Laboratoire Astroparticule Particule Cosmologie, 10, rue Alice Domon et Léonie Duquet
75205 Paris, France}
\begin{abstract}
The high-mass end of the halo mass function is a sensitive probe of primordial non-Gaussianity (NG). In
a recent study \cite{AchitouvCorasaniti2012} we have computed the NG halo mass function in the context of the
Excursion Set theory and shown that the primordial NG imprint is coupled to that induced by
the non-linear collapse of dark matter halos. We also found an excellent agreement with N-body simulation results.
Here, we perform a more accurate computation which accounts for the interval validity of the bispectrum expansion 
to next-to-leading order and extend the calculation to the case of a non-vanishing primordial trispectrum.
\end{abstract} 
\maketitle
\section{Introduction}
The Excursion Set Theory initially introduced by Bond et al. \cite{Bond1991} 
provides a self-consistent mathematical framework to infer the properties of the halo
mass distribution from the statistics of the initial density field.
The formalism generalizes the original Press-Schechter idea \cite{PressSchechter1974} 
by formulating the halo mass counting problem as one of stochastic calculus. The 
starting point is the realization that at any location in space the linear matter density
fluctuation field performs a random walk as function of a filtering scale $R$.
% that
%naturally defines a mass scale $M=\bar{\rho}\,V(R)$,
In average, this scale naturally defines a mass scale $M=\bar{\rho}\,V(R)$,
 where $\bar{\rho}$ is the mean matter 
density and $V(R)$ the enclosed spatial volume. By counting the number of trajectories which first-cross a collapse threshold it is then possible to 
compute the fraction of mass elements in halos $F(M)$ and consequently derive the halo mass function $dn/dM = (1/V)\,dF/dM$.
The requirement of first-crossing is key to solving the so called ``cloud-in-cloud'' 
problem affecting the original Press-Schechter result. In fact, the first-crossing condition guarantees that in the small scale limit 
($R\rightarrow 0$) and independently of the properties of the random walks
the fraction of mass into collapsed objects always tends to unity. 

The Excursion Set is analytically solvable in the case of uncorrelated (Markov) random walks for which the evaluation
of the first-crossing distribution is reduced to solving a standard diffusion problem. However, uncorrelated random walks are generated by
a special filtering of the linear density field which corresponds to a non-physical halo mass definition. In contrast any filtering 
which specifies a physically meaningful mass generates correlated random walks for which $F(M)$ can be inferred only through a numerical computation. 
This has represented a major limitation 
since Monte Carlo simulations are computationally expensive and moreover do not provide the same level of physical insight 
of analytic solutions. The seminal work by Maggiore \& Riotto \cite{MaggioreRiotto2010a} has made a major step forward in this direction. 
Using the path-integral formulation of the Excursion Set Theory the authors have shown that the first-crossing distribution
of correlated random walks can be computed as a perturbative expansion about the Markovian solution.
Using this methodology it has been possible to derive analytical formulae for the halo mass function under different halo collapse model 
assumptions as well as Gaussian and non-Gaussian (NG) initial conditions \cite{MaggioreRiotto2010b,MaggioreRiotto2010c,DeSimone2011}. 

In a series of papers \cite{CorasanitiAchitouv2011a,CorasanitiAchitouv2011b,AchitouvCorasaniti2012}
we have used this formalism to evaluate the imprint of the non-spherical collapse of halos
on the mass function. To this end we have introduced an effective stochastic Diffusive Drifting Barrier (DDB) model which 
parametrizes the main features of the ellipsoidal collapse of halos. 
Accounting for such effects can reproduce the halo mass function from N-body simulations with remarkable 
accuracy both for Gaussian and non-Gaussian initial conditions. 

Here, we extend the work presented \cite{AchitouvCorasaniti2012} to derive a more accurate expression of the
contribution to the halo mass function of the primordial bispectrum expanded in the large scale 
limit to next-to-leading order and compute the leading order contribution of the primordial trispectrum.

The paper is organized as follows. In Section~\ref{secI} we review the path-integral formulation of the Excursion Set and its application to
Gaussian and non-Gaussian initial conditions in the case of the DDB model. In Section~\ref{secII} we evaluate a lower limit on the interval validity
of the bispectrum expansion at next-to-leading order. In Section~\ref{secIII} and~\ref{secIV} compute the bispectrum and trispectrum
contribution to the mass function respectively. Finally, we present our conclusion in Section~\ref{secV}.

\section{Excursion Set Mass Function and Diffusive Drifting Barrier}\label{secI}
Here, we briefly review the main features of the path-integral formulation of the Excursion Set Theory. 
First, let us introduce the variance of the linear density
field $\delta$ filtered on a scale $R$:
\begin{equation}
\sigma^2(R)\equiv S(R)=\frac{1}{2\pi^2}\int dk\,k^2P(k)\tilde{W}^2(k,R),
\end{equation}
where $P(k)$ is the linear matter power spectrum and $\tilde{W}^2(k,R)$ is the Fourier transform of the filter
function in real space, $W(x,R)$. As already mentioned, by selecting a volume $V(R)=\int d^3x\,W(x,R)$ the filter naturally associates a mass to the enclosed region,
$M=\bar{\rho}\,V(R)$. Thus we have a one-to-one relation between $R$ (or $M$) and $S$.

In the Excursion Set Theory the filtered density field $\delta(x,R)$ at any random point in space 
performs a random walk as function of $R$, $M$ or equivalently $S$ which plays
the role of a pseudo-time variable. The random walks start at $S=0$ with $\delta=0$, since in the large scale limit ($R\rightarrow \infty$) we have
$S\rightarrow 0$ and the matter density distribution tends toward homogeneity, i.e. $\delta\rightarrow 0$. We are interested in counting trajectories that at a given value of $S$ 
cross for the first time a collapse density threshold $B$ such that $\delta=B$. This threshold encodes all informations on the gravitational
collapse of halos. In order to model features of the non-spherical collapse, the absorbing barrier $B$ is promoted to a stochastic variable 
(see e.g. \cite{Audit1997,Sheth2001,MaggioreRiotto2010b})
also performing a random walk as function of $S$ (i.e.~$R$ or $M$). In such a case it is convenient to introduce $Y=B-\delta$, which performs
a random walk starting at $Y(0)=Y_0$ with barrier crossing at $Y_c=0$. 

Our goal is to compute the probability distribution $\Pi(Y_0,Y,S)$ 
of trajectories starting at $Y_0$ which reach the value $Y$ at $S$ without ever touching the barrier $Y_c=0$. 
This can be computed as a path-integral over the ensemble of the random trajectories (see \cite{MaggioreRiotto2010a} for a detailed
derivation). Let us discretize the pseudo-time variable $S$ in equally spaced steps, 
$\Delta S=\epsilon$, such that $S_k=k \epsilon$ with $k=1,..,n$. The probability distribution of trajectories starting at $Y_0$ and ending in $Y_n$
at $S_n$ and that have never crossed the barrier before is given by
\begin{equation}
\Pi_\epsilon(Y_0,Y_n,S_n)=\int_{Y_c}^{\infty} dY_1..\int_{Y_c}^{\infty} dY_{n-1} W(Y_0,..,Y_n,S_n),\label{piepsilon}
\end{equation}
where 
\begin{equation}
W(Y_0,..,Y_n,S_n)=\int\mathcal{D}\lambda\, 
  e^{i\sum_{i=1}^n\lambda_i Y_i}\langle e^{-i\sum\limits_{i=1}^n\lambda_i Y(S_i)}\rangle,\label{densitymark}
\end{equation}
where the brackets $\langle...\rangle$ refer to an ensemble average of the random walks
and the averaged quantity is the explonential of
\begin{equation}\label{partition}
Z=\sum_{p=1}^{\infty}\frac{(-i)^p}{p!}\sum_{i_1=1}^{n}...\sum\limits_{i_p=1}^{n}\lambda_{i_1}...\lambda_{i_p}\langle Y_{i_1}..Y_{i_p}\rangle_c
\end{equation}
which is the partition function of the system written in terms of the $p$-point connected correlation functions
$\langle Y_{i_1}..Y_{i_p}\rangle_c$ of the random walks. Thus, the 
the properties of the stochastic system are entirely determined by the connected correlators. Once these are specified
then Eq.~(\ref{piepsilon}) can be integrated in the continuous limit to finally obtain 
the first-crossing distribution
\begin{equation}\label{firstcross}
\frac{dF}{dS}\equiv\mathcal{F}(S)=-\frac{\partial}{\partial S}\left[\int_{Y_c}^{\infty} dY\,\Pi(Y_0,Y,S)\right],
\end{equation}
and the halo mass function is given by
\begin{equation}
\frac{dn}{dM}=f(\sigma)\dfrac{\bar{\rho}}{M^2}\frac{d \log\sigma^{-1}}{d\log M},
\end{equation}
where $f(\sigma)=2S\mathcal{F}(S)$ is the so called ``multiplicity'' function.

\subsection{Gaussian Initial Conditions}
Let us consider a Gaussian density field smoothed with a top-hat filter in real space. On average the density field is homogeneous,
thus implying that $\langle\delta(S)\rangle_c=0$. Therefore, due to the Gaussian nature of the field the only non-vanishing 
connected correlator is the 2-point function $\langle\delta(S)\delta(S')\rangle_c$, while
all higher-order connected correlators identically vanish. Maggiore \& Riotto \cite{MaggioreRiotto2010a} have shown that for 
standard cosmological scenarios with Cold Dark Matter power spectra, the 2-point function smoothed with a sharp-x
is well approximated by $\langle\delta(S)\delta(S')\rangle_c={\rm min}(S,S')+\Delta(S,S')$, where the first term corresponds to Markov random walks
generated by a sharp-k filter and the second term is well approximated by $\Delta(S,S')=\kappa S/S' (S'-S)$ with a nearly constant amplitude $\kappa<1$. 
Thus, the pseudo-time correlations induced by the filter function can be treated as small correction about the Markovian 
case and the mass function obtained using a perturbative expansion of the partition function in the path-integral
in powers of $\kappa$.
 
Concerning the barrier random walks, in \cite{CorasanitiAchitouv2011a,CorasanitiAchitouv2011b} we have introduced
a stochastic model with linear drift and Gaussian diffusion characterized by $\langle B(S)\rangle=\delta_c+\beta S$ and 
$\langle B(S)B(S')\rangle_c=D_B\,{\rm min}(S,S')$, where $\delta_c$ is the linearly extrapolated critical spherical collapse density, $\beta$ is 
the average linear rate of deviation from the spherical collapse prediction and $D_B$ is the amplitude of the scatter about the average 
\footnote{In the Excursion Set the barrier diffusion coefficient parametrizes the stochasticity inherent to the ellipsoidal collapse of halos. However, 
it is important to keep in mind that in the Excursion Set halos can form out of any random position. On the other hand, numerical simulations show
that halos form preferentially out of peaks of the linear density field as suggested by the hierarchical model of structure formation. 
Thus when comparing with N-body results the value of $D_B$ can be biased by the underlying assumption of the Excursion Set approach 
(see \cite{AseemRavi2012} for an extension of the formalism to random walks around density peaks).}. 
In such a case the non-vanishing connected correlators of the $Y$ variable are 
\begin{eqnarray}
\langle Y(S)\rangle_c & = &\delta_c+\beta S \label{yave}\\
\langle Y(S)Y(S')\rangle_c &=& (1+D_B){\rm min}(S,S')+\Delta(S,S').\label{yvar}
\end{eqnarray}
Substituting these expressions in Eq.~(\ref{partition}) and performing a double expansion in $\kappa$ and $\beta$ 
we have derived the Gaussian multiplicity function
\begin{equation}\label{fddb}
f_G(\sigma)=f_{0}(\sigma)+f_{\kappa=1}(\sigma)
\end{equation}
where $f_0(\sigma)$ is the Markovian contribution and $f_{\kappa=1}(\sigma)$ is the filter correction to 
first order in $\kappa$ and up to second order in $\beta$ which read as
\begin{equation}
f_0(\sigma)=\frac{\delta_c}{\sigma}\sqrt{\frac{2a}{\pi}}\,e^{-\frac{a}{2\sigma^2}(\delta_c+\beta\sigma^2)^2}\label{fsigma0}
\end{equation}
and
\begin{equation}\label{fkappa1}
f_{\kappa=1}(\sigma)=f_{1,\beta=0}^{m-m}(\sigma)+f_{1,\beta^{(1)}}^{m-m}(\sigma)+f_{1,\beta^{(2)}}^{m-m}(\sigma)
\end{equation}
with
\begin{equation}
f_{1,\beta=0}^{m-m}(\sigma)=-\tilde{\kappa}\dfrac{\delta_c}{\sigma}\sqrt{\frac{2a}{\pi}}\left[e^{-\frac{a \delta_c^2}{2\sigma^2}}-\frac{1}{2} \Gamma\left(0,\frac{a\delta_C^2}{2\sigma^2}\right)\right],
\end{equation}
\begin{equation}
f_{1,\beta^{(1)}}^{m-m}(\sigma)=- a\,\delta_c\,\beta\left[\tilde{\kappa}\,\text{Erfc}\left( \delta_c\sqrt{\frac{a}{2\sigma^2}}\right)+ f_{1,\beta=0}^{m-m}(\sigma)\right],
\end{equation}
\begin{equation}\label{beta2}
f_{1,\beta^{(2)}}^{m-m}(\sigma)=-a\,\beta\left[\frac{\beta}{2} \sigma^2 f_{1,\beta=0}^{m-m}(\sigma)+\delta_c \,f_{1,\beta^{(1)}}^{m-m}(\sigma)\right],
\end{equation}
where $\tilde{\kappa}=a\,\kappa$ and $a=1/(1+D_B)$. Equation~(\ref{beta2}) includes a term $\mathcal{O}(\beta^2)$ which was missing in the original derivation presented in \cite{CorasanitiAchitouv2011a,CorasanitiAchitouv2011b}. As we explain in Appendix~\ref{app} this is due to having neglected a factor $\exp(-\beta^2S/2)$ in the computation of the probability $\Pi_\epsilon(Y_c,Y_c,S)$\footnote{We thank Ruben van Drongelen for pointing this to us.} and which enters the calculation of the memory-of-memory term to first order in $\kappa$.

\subsection{Non-Gaussian Initial Conditions}
In the case of non-Gaussian initial conditions the higher-order connect correlators of the linear density field are non-vanishing. 
Let us consider the case of primordial non-Gaussianity sourced by a bispectrum term, hence in addition to Eq.~(\ref{yave}) 
and (\ref{yvar}), the partition function contains the contribution of a non-vanishing 3-point connected correlation function 
$\langle Y(S_i)Y(S_j)Y(S_k)\rangle_c=-\langle\delta(S_i)\delta(S_j)\delta(S_k)\rangle_c$ with
\begin{equation}\label{3pts}
\begin{split}
&\langle\delta(S_i)\delta(S_j)\delta(S_k)\rangle_c=\int\dfrac{d^3k_i}{(2\pi)^3}\dfrac{d^3k_j}{(2\pi)^3}\dfrac{d^3k_k}{(2\pi)^3}\tilde{W}(k_i,R_i[S_i])\\
&\times\tilde{W}(k_j,R_j[S_j])\tilde{W}(k_k,R_k[S_k])\mathcal{M}(k_i)\mathcal{M}(k_j)\mathcal{M}(k_j)\times\\
&\times\langle\zeta(\textbf{k}_i)\zeta(\textbf{k}_j)\zeta(\textbf{k}_k)\rangle_c,
\end{split}
\end{equation} 
where $\tilde{W}$ is the Fourier transform of the sharp-x filter, $\mathcal{M}(k)=2/(5 H_{0}^{2}\Omega_m)T(k)k^2$, 
$H_0$ is the Hubble constant, $\Omega_m$ the matter density, $T(k)$ the transfer function and $\zeta(k)$ 
is the curvature perturbation with
\begin{equation}
\langle\zeta(\textbf{k}_i)\zeta(\textbf{k}_j)\zeta(\textbf{k}_k)\rangle_c=(2\pi)^3\delta_D(\textbf{k}_i+\textbf{k}_j+\textbf{k}_k) B(k_i,k_j,k_k),
\end{equation}
where $B(k_i,k_j,k_k)$ is the so called ``reduced'' bispectrum. 

By expanding Eq.~(\ref{densitymark}) in powers of the amplitude of the reduced bispectrum (usually parametrized by the coefficient $f_{NL}$),
we obtain to first-order in $f_{NL}$ the non-Gaussian part of the first-crossing distribution
\begin{equation}
\mathcal{F}_{NG}(S)=-\frac{\partial}{\partial S} F_{NG}(S)
\end{equation}
where $F_{NG}(S)$ is the continuous limit of
\begin{equation}
\begin{split}\label{Fng}
&F_{NG}(S)=\frac{1}{6}\sum_{i,j,k=0}^{n}\langle\delta(S_i)\delta(S_j)\delta(S_k)\rangle_c\times\\
&\times\int_{Y_c}^{\infty}dY\int_{Y_c}^{\infty}dY_1...dY_{n-1}\partial_i\partial_j\partial_k W_{0}(Y_0,...,Y,S),
\end{split}
\end{equation}
where $W_0(...)$ is the Gaussian Markovian probability density distribution.
Eq.~(\ref{Fng}) can be evaluated provided we have an analytical expression for the primordial bispectrum. 
In \cite{AchitouvCorasaniti2012} we have used the standard approach of considering a triple Taylor series 
of the primordial bispectrum in the large scale limit. 
In the next Section we will study in detail the range of validity of such an expansion and infer the
relevant contribution to the non-Gaussian halo mass function.

\section{Interval Validity of Primordial Bispectrum Expansion}\label{secII}
Let us expand the bispectrum Eq.~(\ref{3pts}) in a triple Taylor series around $S_i\sim S_j\sim S_k\sim S$ (for convenience we set $S=S_n$):
\begin{equation}\label{Som}
\begin{split}
&\langle\delta(S_i)\delta(S_j)\delta(S_k)\rangle_c =\sum_{p,q,r=0}^{\infty}\dfrac{(-1)^{p+q+r}}{p!q!r!}
(S-S_i)^p\times\\&\times(S-S_i)^q(S-S_k)^r G_{3}^{(p,q,r)}(S)
\end{split}
\end{equation}
where 
\begin{equation}
G_{3}^{(p,q,r)}(S)\equiv\frac{d^p}{dS_{i}^{p}}\frac{d^q}{dS_{j}^{q}}\frac{d^r}{dS_{k}^{r}} \langle\delta(S_i)\delta(S_j)\delta(S_k)\rangle_c\bigg|_{i,j,k=n}.
\end{equation}

We expect the signature of primordial non-Gaussianity to be stronger at large scales ($S\rightarrow 0$) where the evolution
of the density field remains linear. Hence, the leading order contribution to the halo mass function is given by 
the lowest order term of the bispectrum expansion. This corresponds to having $p+q+r=0$ which gives the leading order 
term $\langle\delta^3(S)\rangle$ that can be computed numerically using Eq.~(\ref{3pts}) for a given
type of primordial NG. It is convenient to introduce the normalized skewness $\mathcal{S}_3(S)=\langle\delta^3(S)\rangle/S^2$. 
In \cite{AchitouvCorasaniti2012} we have provided fitting formula
for $\mathcal{S}_3(S)$ and its derivatives accurate to a few percent for local and equilateral non-Gaussianity.
 
The next-to-leading order contribution is given by three terms corresponding
to the case $p+q+r=1$. Hence, up to next-to-leading order the primordial bispectrum reads as
\begin{equation}\label{expg3}
\begin{split}
&\langle\delta(S_i)\delta(S_j)\delta(S_k)\rangle_c =\langle\delta^3(S)\rangle-(S-S_i)\,G_{3}^{(1,0,0)}(S)+\\&-(S-S_j)\,G_{3}^{(0,1,0)}(S)-(S-S_k)\,G_{3}^{(0,0,1)}(S),
\end{split}
\end{equation}
since $S_i\sim S_j\sim S_k$ we can collect the terms in $(S-S_i)$, moreover by computing $G_3^{(1,0,0)}$, $G_3^{(0,1,0)}$ and $G_3^{(0,0,1)}$ as
derivatives of Eq.~(\ref{3pts}) one can notice that 
\begin{equation}\label{sumg3}
G_{3}^{1,0,0}(S)+G_{3}^{0,1,0}(S)+G_{3}^{0,0,1}(S)=\dfrac{dR}{dS}\dfrac{d}{dR}\langle\delta^3(R)\rangle
\end{equation}
and introducing 
\begin{equation}
\mathcal{U}_3(S)=\dfrac{1}{S}\dfrac{dR}{dS}\dfrac{d}{dR}\langle\delta^3(R(S))\rangle,
\end{equation}
we can rewrite Eq.~(\ref{expg3}) as
\begin{equation}\label{bispdec}
\langle\delta(S_i)\delta(S_j)\delta(S_k)\rangle_c\approx S^2\mathcal{S}_3(S) - (S-S_i) S\,\mathcal{U}_3(S).
\end{equation}
We can now derive a lower limit, $S_{\rm min}$, on the value of $S_i,S_j,S_k$ for which such an expansion remains valid.
This is obtained by imposing the next-to-leading order term to be smaller than the leading one.
We find 
\begin{equation}
S_{\rm min}=S\left[1-\frac{\mathcal{S}_3(S)}{\mathcal{U}_3(S)}\right],
\end{equation}
using the fitting formulae for $\mathcal{S}_3(S)$ and $\mathcal{U}_3(S)$ derived in \cite{AchitouvCorasaniti2012} we find that
to good approximation $S_{\rm min}=S/\alpha$ with $\alpha^{-1}_{\rm loc}= 0.373$ and $\alpha^{-1}_{\rm equi}= 0.382$ 
for local and equilateral NG respectively.

\section{Non-Gaussian Halo Mass Function and Bispectrum Expansion Accuracy}\label{secIII}
Having inferred a lower limit on the interval validity of the bispectrum expansion up to next-to-leading
order we can infer a more accurate estimate of its contribution to the multiplicity function.
Given the bispectrum expansion Eq.~(\ref{bispdec}) we can split Eq.~(\ref{Fng}) as
\begin{equation}
F_{NG}(S)=F_{NG}^L(S)+F_{NG}^{NL}(S).
\end{equation}
As shown in \cite{AchitouvCorasaniti2012} the leading order term is given by 
\begin{equation}
F_{NG}^L(S)=\frac{1}{6}S^2 S_3 (S)\int_{Y_c}^{\infty}dY\,\frac{\partial^3}{\partial Y_{c}^3} \Pi_0(Y_0,Y,S)
\end{equation}
where $\Pi_0(Y_0,Y,S)$ is the probability distribution of the Gaussian random walks in the case of the Diffusing Drifting Barrier model given by Eq.~(8) in \cite{CorasanitiAchitouv2011a}. The integral can be computed analytically to finally obtain the 
leading order contribution to the multiplicity function \cite{AchitouvCorasaniti2012}:
\begin{widetext}
\begin{equation}\label{fngl}
\begin{split}
f_{NG}^{L}(\sigma)=\frac{a}{6}\sqrt{\frac{2a}{\pi}}\sigma e^{-\dfrac{a(\delta_c+\beta\sigma^2)^2}{2\sigma^2}}\biggl\{S_3(\sigma)\biggl[\dfrac{a^2}{\sigma^4}\delta_c^4
-2\dfrac{a}{\sigma^2}\delta_c^2-1+3\dfrac{a^2}{\sigma^2}\beta\delta_c^3+3 a\beta \delta_c+a^2\beta^3\sigma^2 \delta_c+3 a^2\beta^2\delta_c^2+13 a \beta^2\sigma^2\biggr]+\\
+\dfrac{dS_3(\sigma)}{d\log\sigma}\biggl[\dfrac{a}{\sigma^2}\delta_c^2-1+3 a\beta \delta_c+4 a\beta^2\sigma^2\biggr] \biggr\}
+\dfrac{2}{3}a^3\beta^3\sigma^4 e^{-2a\beta \delta_c}\text{Erfc}\biggl[\sqrt{\dfrac{a}{2\sigma^2}}(\delta_c-\beta\sigma^2)\biggr]\biggl\{ 4 S_3(\sigma)+\dfrac{dS_3(\sigma)}{d\log\sigma}\biggr\} 
\end{split}
\end{equation}
\end{widetext}

On the other hand let us detail more the derivation of the next-to-leading order term which differs from
that of \cite{AchitouvCorasaniti2012}. In such a case we have from Eq.~(\ref{Fng})
that
\begin{equation}
\begin{split}
F_{NG}^{NL}(S)&=-\frac{1}{6}S\,U_3 (S)\sum_{i_{\rm{min}}}^{i_{\rm{max}}}(S-S_i)\times\\
&\times\int_{Y_c}^{\infty} dY\sum_{j,k=1}^{n}\int_{Y_c}^{\infty}dY_1...\int_{Y_c}^{\infty} dY_{n-1} \,\partial_i\partial_j\partial_k W_0,
\end{split}
\end{equation}
where we have decomposed the sum in Eq.~(\ref{Fng}) and kept only the terms up to $n-1$. In fact, as shown in 
\cite{AchitouvCorasaniti2012} the integral in $dY_n$ of the terms with $i,j,k=n$ 
vanishes since the integrands are total derivatives. Furthermore it is easy to show that 
$\sum_{j,k} \rightarrow {\partial^2}/{\partial Y_c^2}$, thus
\begin{equation}
\begin{split}
F_{NG}^{NL}(S)&=-\frac{1}{6}S\,U_3 (S)\sum_{i=i_{\rm{min}}}^{n-1}(S-S_i)\times\\
&\times\int_{Y_c}^{\infty} dY\frac{\partial^2}{\partial Y_c^2} \biggl[ \int_{Y_c}^{\infty}dY_1...\int_{Y_c}^{\infty} dY_{n-1} \,\partial_i W_0\biggr],
\end{split}
\end{equation}
where the sum is bounded from below due to the fact that $S_{min}<S_i<S$.
The multiple integral in the above expression can be computed by part and using the fact that $W_0$ obeys the Chapman-Kolmogorov
equation 
\begin{equation}
\begin{split}
&W_0(Y_0,..,\hat{Y}_i,..,Y,S)=W_0(Y_0,..,Y_c,S_i-S_{i_{\rm{min}}})\times\\
&\times W_0(Y_c,..,Y,S-S_i+S_{i_{\rm{min}}})
\end{split}
\end{equation}
we have
\begin{equation}
\begin{split}
&F_{NG}^{NL}=-\frac{1}{6}S\,U_3 (S)\sum_{i=i_{\rm{min}}}^{n-1}(S-S_i)\times\\
&\times\int_{Y_c}^{\infty}dY\frac{\partial^2}{\partial Y_c^2}\biggl[\Pi\biggl(Y_0,Y_c,S_i-\frac{S}{\alpha}\biggr)\Pi\biggl(Y_c,Y,S-S_i+\frac{S}{\alpha}\biggr)\biggr].
 \end{split}
\end{equation}
where $\Pi(Y_0,Y_c,S_i-S/\alpha)$ and $\Pi(Y_c,Y,S-S_i+S/\alpha)$ are given by Eq.~(3.20) and (3.21) in \cite{AchitouvCorasaniti2012}
respectively. In the continuous limit and taking into account that the bispectrum expansion up to next-to-leading order is valid in the range
 $S_{min}\approx S/\alpha<S_i<S$ we have $\sum_{i=i_{\rm{min}}}^{n-1}\rightarrow \int_{S/\alpha}^{S}dS_i$. 

Finally, we 
obtain the first-crossing distribution to next-to-leading order which reads as
\begin{equation}
\mathcal{F}_{NG}^{NL}(S)=-\frac{1}{6}\left\{\frac{\partial^2 I_{NL}^{\alpha}}{dY_{c}^2}\left[U_3+S\frac{dU_3}{dS}\right]+S U_3\frac{\partial}{\partial S}\frac{\partial^2 I_{NL}^{\alpha}}{dY_{c}^2}\right\}
\end{equation}
with 
\begin{equation}\label{d2idyc2}
\begin{split}
&\frac{\partial^2 I_{NL}^{\alpha}}{\partial Y_{c}^2}=-\frac{a}{\pi}\int_{\frac{S}{\alpha}}^{S}\frac{dS_i}{S_{i}^{3/2}}e^{-\frac{a}{2S_i}(\delta_c+\beta S_i)^2}\left[S\left(1-\frac{1}{\alpha}\right)-S_i\right]\times\\
&\times\left\{-2 a \beta+a^2\beta^2 \delta_c+a\frac{\delta_c}{S_i}\left[-3+2 a \beta \delta_c+a\frac{\delta_c^2}{S_i}\right]\right\}\times\\
&\times\left\{e^{-\frac{a}{2}\beta^2(S-S_i)}+\frac{\beta}{2}\sqrt{\pi\,a(S-S_i)}\text{Erfc}\left[-\beta\sqrt{\frac{a}{2}(S-S_i)}\right]\right\}
\end{split}
\end{equation}
and
\begin{equation}\label{ddsd2idyc2}
\begin{split}
&\frac{\partial}{\partial S}\frac{\partial^2 I_{NL}^{\alpha}}{\partial Y_{c}^2}=-\frac{a}{\pi}\int_{\frac{S}{\alpha}}^{S}\frac{dS_i}{S_{i}^{3/2}} e^{-\frac{a}{2S_i}(Y_0+\beta S_i)^2}\times\\
&\times\left\{-2 a \beta+a^2\beta^2 \delta_c+a\frac{\delta_c}{S_i}\left[-3+2 a \beta \delta_c+a\frac{\delta_c^2}{S_i}\right]\right\}\times\\
&\times\left\{\frac{e^{-\frac{a}{2}\beta^2(S-S_i)}}{2\sqrt{S-S_i}}\left[1-\frac{1}{\alpha}+\frac{1}{\alpha}\frac{S_i}{S-S_i}+\frac{1}{\alpha}a\beta^2\frac{S+S_i}{\sqrt{S-S_i}}\right]+\right.\\
&\left.+\sqrt{\frac{a\pi}{2}}\beta\left(1-\frac{1}{\alpha}\right)\text{Erfc}\left[-\beta\sqrt{\frac{a}{2}(S-S_i)}\right]\right\}.
\end{split}
\end{equation}

\begin{figure}[ht]
\centering
\begin{tabular}{cc}
\includegraphics[scale=0.35]{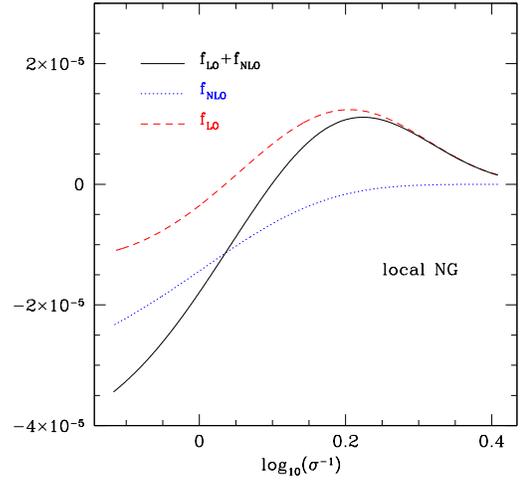}\\
\includegraphics[scale=0.35]{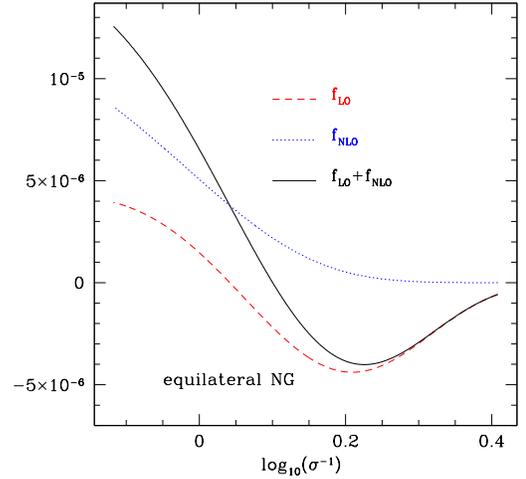}
\end{tabular}
\caption{Leading order (red dash line), next-to-leading order (blue dot line) contribution to the non-Gaussian multiplicity function (black solid line) 
for local (top panel) and equilateral (bottom panel) non-Gaussianities respectively.}\label{fig1}
\end{figure}

The integrals in Eq.~(\ref{d2idyc2}) and (\ref{ddsd2idyc2}) can be computed numerically to finally evaluate the next-to-leading order contribution to the 
multiplicity function, 
%\begin{equation}
$f_{NG}^{NL}(\sigma)=2\sigma^2 \mathcal{F}_{NG}^{NL}(\sigma^2)$.
%\end{equation}

In Fig.~\ref{fig1} we plot $f_{NG}^{L}(\sigma)$, $f_{NG}^{NL}(\sigma)$ and their sum in units of $f_{NL}$
for local (top panel) and equilateral (bottom panel) non-Gaussianity. We have set the
DDB model parameters to the values best fitting the Gaussian halo mass function inferred from Gaussian N-body simulations in \cite{PPH8}. 
The mass interval shown here corresponds to that probed by these numerical simulations. We may notice that in the local and equilateral cases the leading
order term is larger than the next-to-leading order one at high-masses ($\log_{10}\sigma^{-1}>0.1$), while at lower masses the next-to-leading order is larger. 
This is expected since as already mentioned the signature of primordial non-Gaussianity at large masses results of
the lowest order in the bispectrum expansion. 

\begin{figure}[ht]
\centering
\begin{tabular}{cc}
\includegraphics[scale=0.35]{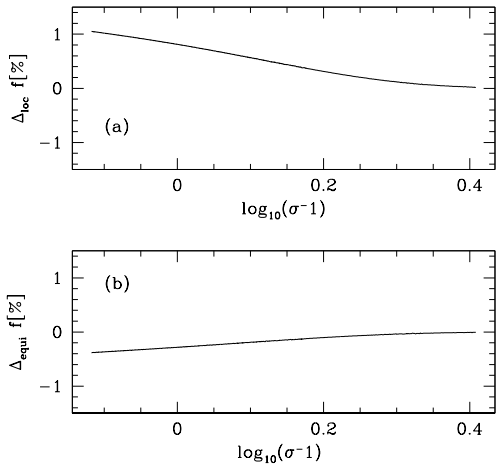}
\end{tabular}
\caption{Relative difference bewteen of the non-Gaussian halo mass function with and without next-leading order contribution
in the case of local (panel a) and equilateral non-Gaussianity (panel b).}\label{fig3}
\end{figure}

We can now evaluate the overall contribution to the halo multiplicity function, $f(\sigma)=f_G(\sigma)+f_{NG}(\sigma)$.
In Fig.~\ref{fig3} we plot the relative difference of the NG halo mass function  with and without next-to-leading order term for local (top panel)
and (bottom panel) equilateral non-Gaussianity respectively in the case of $f_{NL}=150$. As we can see the differences is no larger than $2\%$
in the low mass range, hence the next-to-leading term remains negligible even for large non-Gaussianities in the mass range
corresponding to halos with $M>10^{13} M_\odot$ and can be neglected for practical purposes.

\section{Trispectrum contribution to the Non-Gaussian Halo Mass Function}\label{secIV}
A number of scenarios of primordial inflation predict deviation from Gaussianity of the form (see e.g. \cite{Sasaki2006,Enqvist2008})
\begin{equation}\label{zeta4}
\zeta=\zeta_G+\frac{3}{5}f_{NL}(\zeta_{G}^2-\langle\zeta_{G}^2\rangle)+\frac{9}{25}g_{NL}\zeta_{G}^3+\mathcal{O}(\zeta_{G}^{4}),
\end{equation}
where $\zeta_{G}$ is the primordial curvature perturbation and $g_{NL}$ is the amplitude of the cubic term which give rises to
a non-vanishing 4-point connected correlation function of the linear density field given by
\begin{equation}\label{4pts}
\begin{split}
&\langle\delta(S_i)\delta(S_j)\delta(S_k)\delta(S_l)\rangle_c=\int\dfrac{d^3k_i}{(2\pi)^3}\dfrac{d^3k_j}{(2\pi)^3}\dfrac{d^3k_k}{(2\pi)^3} \dfrac{d^3k_l}{(2\pi)^3}\\
&\tilde{W}(k_i,R_i)\tilde{W}(k_j,R_j)\tilde{W}(k_k,R_k)\tilde{W}(k_l,R_l)\times\\
&\times\mathcal{M}(k_i)\mathcal{M}(k_j)\mathcal{M}(k_k)\mathcal{M}(k_l)\langle\zeta(\textbf{k}_i)\zeta(\textbf{k}_j)\zeta(\textbf{k}_k)\zeta(\textbf{k}_l)\rangle_c
\end{split}
\end{equation}
where $\langle\zeta(\textbf{k}_i)\zeta(\textbf{k}_j)\zeta(\textbf{k}_k)\zeta(\textbf{k}_l)\rangle_c=(2\pi)^3\delta_D(\textbf{k}_i+\textbf{k}_j+\textbf{k}_k+\textbf{k}_l) T(k_i,k_j,k_k,k_l)$
and $T(k_i,k_j,k_k,k_l)$ is the trispectrum. 

We can compute the trispectrum contribution to the multiplicity function by including the 4-point connect correlator
in the partition function and expand the path-integral for small values of the trispectrum amplitude. As in the case of
the bispectrum, to leading order in a large scale expansion the trispectrum can be approximated as
\begin{equation}
\langle\delta(S_i)\delta(S_j)\delta(S_k)\delta(S_l)\rangle_c\simeq\langle\delta^4(S)\rangle_c
\end{equation}
To first-order in the trispectrum amplitude we have
\begin{equation}\label{fcrstri}
\mathcal{F}_{NG}^{\rm Tri,L}(S)=-\dfrac{\partial}{\partial S}F_{NG}^{\rm Tri,L}(S),
\end{equation}
where $F_{NG}^{\rm Tri,L}(S)$ is the continuous limit of 
\begin{equation}
\begin{split}
&F_{NG}^{\rm Tri,L}(S)=-\dfrac{1}{4!}\sum_{i,j,k,l}\langle\delta^4(S)\rangle_c\times\\
&\times\int_{Y_c}^{\infty}dY \int_{Y_c}^{\infty}dY_1...dY_{n-1}\partial_i\partial_j\partial_k\partial_l W_0.
\end{split}
\end{equation}
Using the fact that $\sum_{i,j,k,l} \rightarrow {\partial^4}/{\partial Y_c^4}$ we obtain
\begin{equation}\label{fngtri}
F_{NG}^{\rm Tri,L}(S)=-\dfrac{1}{4!}\langle\delta^4(S)\rangle_c\frac{\partial^4}{\partial Y_{c}^{4}}\int_{Y_c}^{\infty}\Pi_0(Y_0,Y,S)\,dY,
\end{equation}
the integral can be computed analytically, 
\begin{equation}
\begin{split}
&\frac{\partial^4}{\partial Y_{c}^{4}}\int_{Y_c}^{\infty} \Pi(Y_0,Y,S)\,dY=\sqrt{\frac{a}{2\pi S}}e^{\frac{a}{2S}(\delta_c+\beta S)^2}\times\\
&\times\biggl[-16(a\beta)^3+8\frac{a^2}{S}\beta+6\frac{a^2}{S^2}Y_0-14\frac{a^3}{S}\beta^2 \delta_c-8\frac{a^3}{S^2}\beta \delta_c^{2}+\\
&-2\frac{a^3}{S^3}\delta_c^3\biggr]-8(a\beta)^4e^{-2a\beta Y_0}\text{Erfc}\left[\sqrt{\frac{a}{2S}}(Y_0-\beta S)\right].
\end{split}
\end{equation}

As in the case of the bispectrum, it is convenient to introduce the 4th-order reduced cumulant,
%\begin{equation}
%$\mathcal{S}_{4}(R)\equiv\frac{\langle\delta^4(R)\rangle}{S^3}$
$\mathcal{S}_{4}(R)\equiv\langle\delta^4(R)\rangle/S^3$.
%\end{equation}
Substituting in Eq.~(\ref{fngtri}) and evaluating the first-crossing distribution Eq.~(\ref{fcrstri})
we finally obtain the trispectrum contribution to the multiplicity function:
\begin{equation}
\begin{split}\label{fL4}
&f_{NG}^{\rm Tri,L}(\sigma)=2\,\mathcal{S}_4(\sigma)\sigma^6 (a\beta)^4 e^{-2a\beta \delta_c}\times\\
&\times\text{Erfc}\left[\sqrt{\frac{a}{2\sigma^2}}(\delta_c-\beta\sigma^2)\right]+\mathcal{S}_4(\sigma)e^{-\frac{a}{2\sigma^2}(\delta_c+\beta\sigma^2)^2}\times\\
&\times(a\sigma)^2\sqrt{\dfrac{2a}{\pi S}}\biggl[-\frac{1}{2}\beta\sigma^2+\frac{11}{6}a\beta^3\sigma^4-\frac{1}{8}\delta_c+a(\beta\sigma)^2\delta_c+\\
&+\frac{1}{24}a^2(\beta\sigma)^4\delta_c+\frac{1}{6}(a\sigma \delta_c)^2\beta^3+\frac{1}{4}(a\beta)^2\delta_c^3-\frac{1}{6}a\frac{\delta_c^3}{\sigma^2}+\\
&+\frac{1}{6}\frac{a^2}{\sigma^2}\beta\delta_c^4+\frac{1}{24}a^2\frac{\delta_c^5}{\sigma^4}\biggr]+\frac{1}{3}\frac{d\mathcal{S}_4(\sigma)}{d\log\sigma}\sigma^6 (a\beta)^4 e^{-2a\beta \delta_c}\times\\
&\times\text{Erfc}\left[\sqrt{\frac{a}{2\sigma^2}}(\delta_c-\beta\sigma^2)\right]+\frac{d\mathcal{S}_4(\sigma)}{d\log\sigma}e^{-\frac{a}{2\sigma^2}(\delta_c+\beta\sigma^2)^2}\times\\
&\times(a\sigma)^2\sqrt{\dfrac{2a}{\pi S}}\biggl[-\frac{1}{6}\beta\sigma^2+\frac{1}{3}a\beta^3\sigma^4-\frac{1}{8}\delta_c+\\
&+\frac{7}{24}a(\beta\sigma)^2 \delta_c+\frac{1}{6}a\beta \delta_c^2+\frac{1}{24}a\frac{\delta_c^3}{\sigma^2}\biggr]
\end{split}
\end{equation}

It can be noticed that by setting the barrier model parameters to the spherical collapse values $a=1$ and $\beta=0$ we recover the formula
derived in \cite{MaggioreRiotto2010d}. It is also worth noticing that in the spherical collapse limit and 
neglecting the filter correction to first order in $\kappa$, the NG multiplicity function given by the sum of the Markovian term, the bispectrum and trispectrum leading order contributions has the same functional form as that derived in \cite{LV} using the Edgeworth expansion to describe the non-Gaussian probability distribution of the initial density perturbations.

%\textbf{
The above formula has been derived without making any assumption on the mechanism that generates the non-vanishing 4-point correlation function
of the primordial density field, namely the specific form of the trispectrum, $T(k_i,k_j,k_k,k_l)$. Furthermore, the  
amplitude of the trispectrum is affected not only by the cubic term in 
Eq.~(\ref{zeta4}) and parametrized in terms of $g_{NL}$, but also by the skewness which is parametrized by $f_{NL}$. 
In models where curvature perturbations are sourced by a single scalar field, as in the case of the curvaton model 
(see e.g. \cite{LythWands2002}) the skewness contribution to the kurtois (parametrized in terms of $\tau_{NL}$) 
is given by $\tau_{NL}=36/25 f_{NL}^2$. In \cite{SuyamaYamaguci2008} it has been shown that for a variety
of inflationary scenarios holds the disequality $\tau_{NL}\ge 36/25 f_{NL}^2$. Recently, 
the authors of \cite{Biagettietal2012} have argued that violating such an inequality would imply
some non-trivial new physics, since the inequality results on the one hand from the fact that NG is generated on super-horizon scales 
and on the other hand on the positivity of the 2-point correlation function. Thus, testing such inequality may provide 
hints of fundamental physics at the epoch of inflation.
%}

%\textbf{
It is beyond the scope of this paper to compute the trispectrum
for specific primordial non-Gaussian scenarios for which the forth-order reduced cumulant needs
to be numerically computed for a given trispectrum template. Hence, for simplicity we limit to local type of non-Gaussianity for
fitting functions of $\mathcal{S}_4(\sigma)$ has been computed in \cite{Yokoyama2011}. Even in such a restricted case
we can still infer some relevant information on the imprint of the trispectrum on the halo mass function and the implication
for testing the Suyama-Yamaguchi inequality.
%}

%\textbf{
\begin{figure}[ht]
\centering
\begin{tabular}{cc}
\includegraphics[scale=0.35]{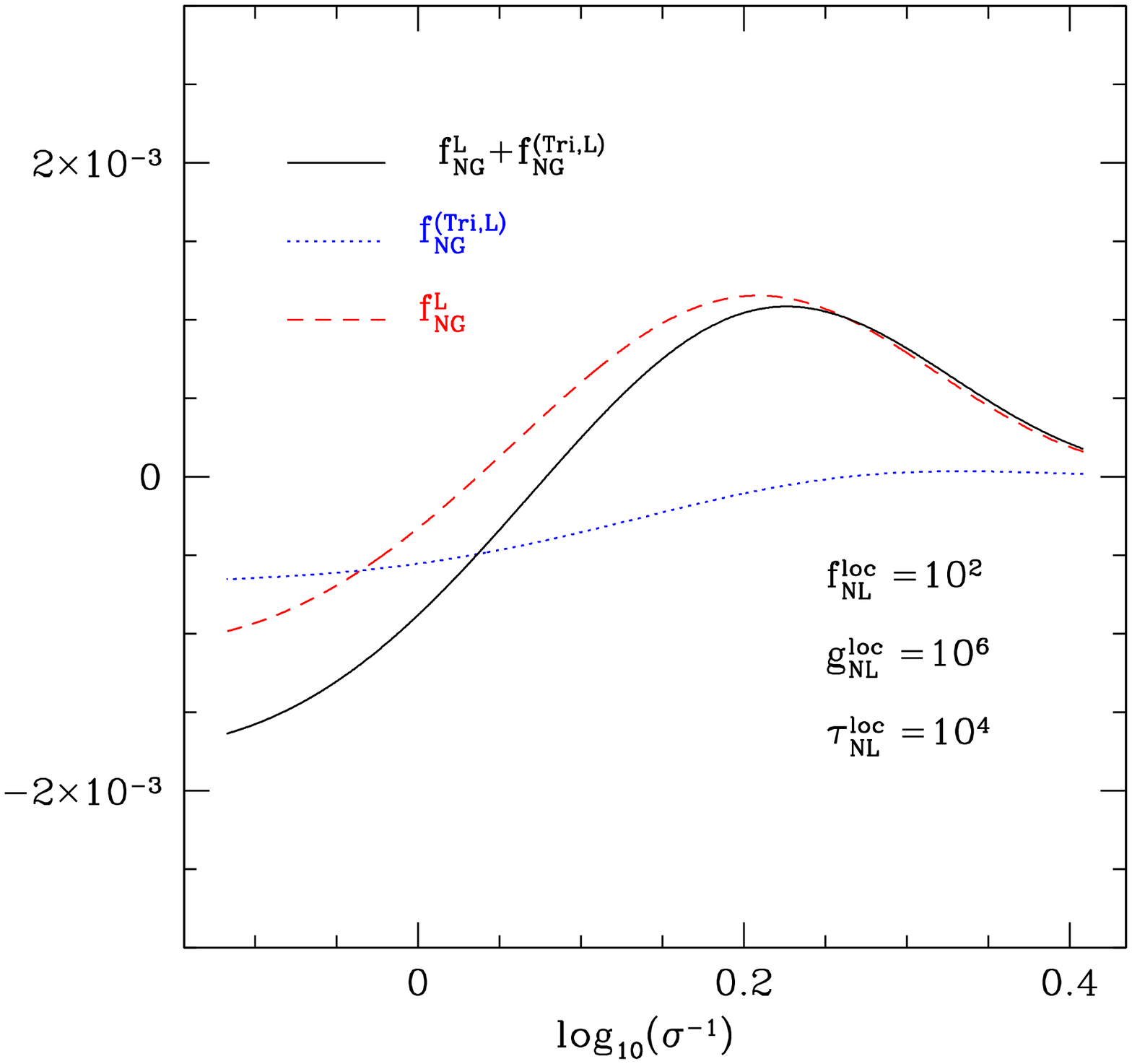}\\
\includegraphics[scale=0.35]{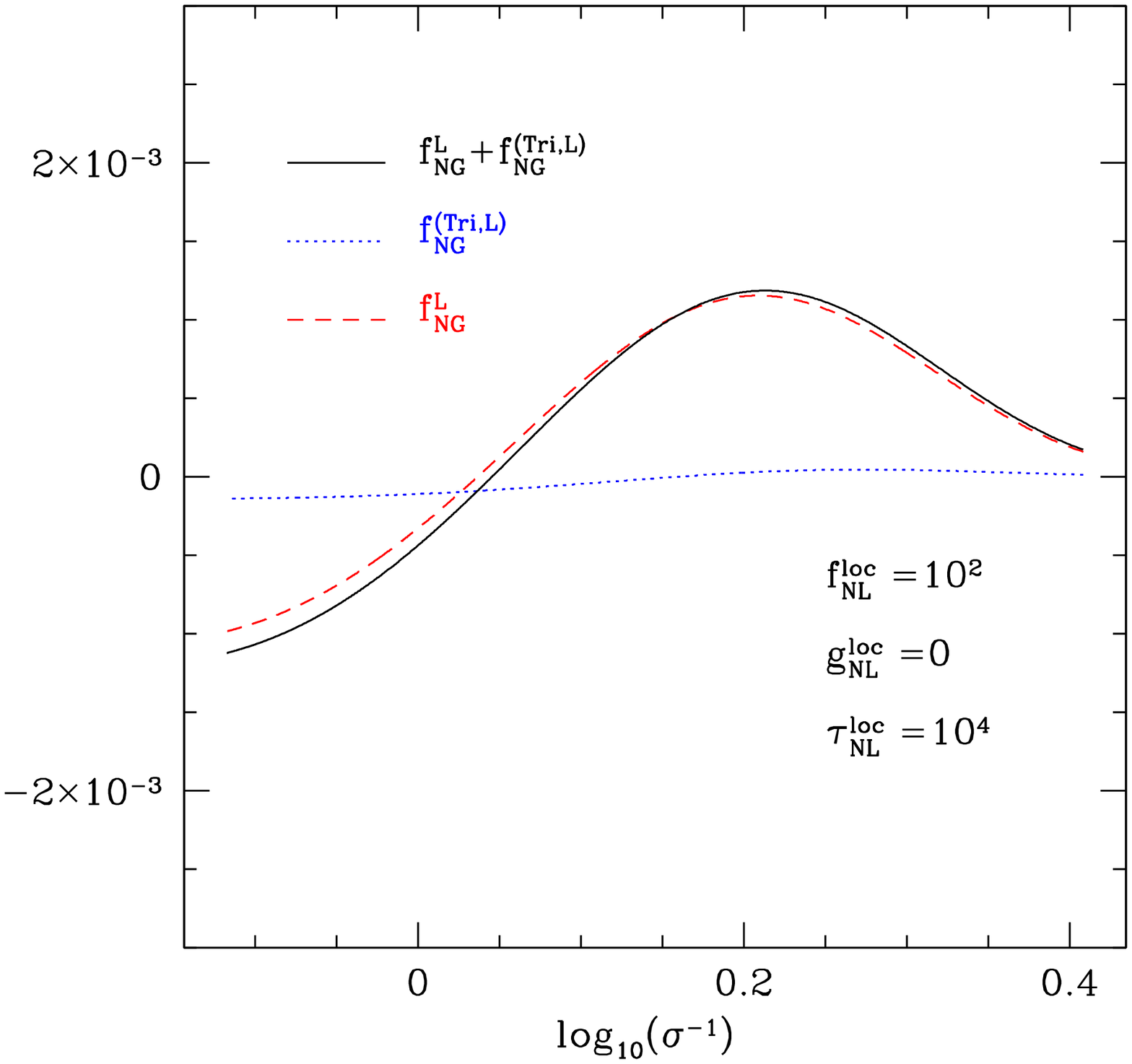}
\end{tabular}
\caption{Leading order contribution of the bispectrum (red dash line), the trispectrum (blue dot line) 
and their sum (black solid line) for local type of primordial non-Gaussianity for $f_{NL}=100$, $\tau_{NL}=10^4$ in the case of 
$g_{NL}=0$ (bottom panel) and $g_{NL}=10^6$ (top panel) respectively.}\label{fig4}
\end{figure}
To this purpose we set the DDB model parameters to their Gaussian value\footnote{As shown in \cite{AchitouvCorasaniti2012} 
this is a good approxiamtion for $f_{NL}<150$.} and plot in Fig.~\ref{fig4}
the contribution of the bispectrum (red dash line), trispectrum (blue dot line) and their sum (black solid line) to the multiplicity function
for values of primordial non-Gaussian amplitude which are consistent with Cosmic Microwave Background (CMB) limits \cite{Smidt2010}. 
In particular, we set $f_{NL}=100$, $\tau_{NL}=10^4$ and consider $g_{NL}=0$ (bottom panel) and $g_{NL}=10^6$ (top panel) respectively. 
We can see that even for $g_{NL}=10^6$ and $\tau_{NL}=10^4$ the trispectrum signal exceeds that of the bispectrum 
only in a very limited low-mass range. In contrast, for $g_{NL}=0$ and $\tau_{NL}=10^4$ the trispectrum  
contribution to the non-Gaussian multiplicity function remains very small compared to that of the bispectrum.
This suggests that within current CMB limits, a violation of the Suyama-Yamaguci inequality
will be hardly detectable solely using the halo mass function. It is possible that measurements of the galaxy bias
may be more informative as investigated in \cite{Biagettietal2012}. On the other hand constraints on the halo mass function
from cluster counts may still be a useful probe when used in combination with estimates of the halo bias 
from the clustering of massive clusters. As shown in \cite{Pillepich2012} for the case of primordial bispectrum, these tests can provide 
improved constraints on primordial non-Gaussianity from the upcoming generation of cosmic structure surveys.
%} 

\section{Conclusion}\label{secV}
The halo mass function carries an imprint of the statistics of the primordial density field as well as the properties of the halo
collapse process. The path-integral formulation of the Excursion Set theory provides a powerful and self-consistent mathematical
framework to account for these effects on the halo mass function. Here, we have extended a previous analysis \cite{AchitouvCorasaniti2012} and 
performed a more accurate derivation of the contribution of the primordial bispectrum expanded in the large
scale limit to next-to-leading order for the Diffusive Drifting Barrier model introduced 
in \cite{CorasanitiAchitouv2011a,CorasanitiAchitouv2011b}. We have shown that 
the next-to-leading order term of the primordial bispectrum decomposition contributes to no more than $\sim2\%$ of the 
non-Gaussian mass function. Thus, for all practical purposes it can be neglected. We have also derived an analytic formula
for the trispectrum contribution. As in the case of the bispectrum, the multiplicity function depends on terms which couple the 
parameters encoding the ellipsoidal collapse of halos with the primordial four-point correlation function.
Also in this case we find that in the spherical collapse limit the trispectrum contribution reduces to the functional 
form derived in the Press-Schechter formalism using the Edgeworth expansion. However, in order to reproduce N-body simulation results the
latter requires two ad-hoc prescriptions. First, the non-Gaussian prediction is rescaled by a Gaussian simulation 
calibrated multiplicity function, such as to account for the imprints of the ellipsoidal collapse. Thus, implicitly assuming that the effect of the non-spherical collapse of halos on the mass function is independent of the amplitude of primordial non-Gaussianity. A good modelling of the collapse parameters is even more relevant for primordial trispectrum terms which act at low masses where the simple spherical collapse is not valid since qualitatively the trispectrum signature on the halo mass function can be mimic by a higher value of $\beta$. In principle, the later can probe multi-field inflation by testing the validity of the Suyama-Yamaguchi inequality. However, assuming values of $f_{NL}$ and $\tau_{NL}$ consistent with current CMB limits, we find that bispectrum contribution is always the dominant NG signal. It is possible that tests of the scale dependent halo bias can be more informative regarding the trispectrum signature. In such a case, it will be interesting to investigate, in the context of the peak background split,  how the mass filtering corrections and the coupling between non-spherical collapse parameters and primordial non-Gaussian amplitudes alter the linear halo bias prediction.

\begin{acknowledgments}
We thank James G. Bartlett for his support and his advices. We also thank Ruben van Drongelen and Koenraad Schalm for useful discussions. I. Achitouv is supported by a scholarship of the `Minist\`ere de l'Education Nationale, de la Recherche et de la Technologie' (MENRT). The research leading to these results has received funding from the European Research Council under the European Community's Seventh Framework Programme (FP7/2007-2013 Grant Agreement no. 279954).
\end{acknowledgments}

\appendix
\section{}\label{app}
Here, we derive the finite correction to the probability distribution of the random walks starting and ending at the barrier,
$\Pi_{\varepsilon}(Y_c,Y_c,S_n)$ which enters the calculation of the memory-of-memory term due to the non-Markovian filter corrections
(see Eq.~(11) in \cite{CorasanitiAchitouv2011a}).
Following the derivation presented in \cite{CorasanitiAchitouv2011b} let us consider the Chapman-Kolmogorov equation:
\begin{equation}\label{chapkol}
\begin{split}
&\Pi_{\varepsilon}(Y_0,Y_n,S_n)=\int_{Y_c}^\infty dY_{n-1}\psi_{\varepsilon}(\Delta Y-\beta\varepsilon)\times\\
&\times\Pi_{\varepsilon}(Y_0,Y_{n-1},S_{n-1}),
\end{split}
\end{equation}
with 
\begin{equation}
\psi_{\varepsilon}(\Delta Y-\beta\varepsilon)=\sqrt{\frac{a}{2\pi\varepsilon}}e^{-a\frac{(\Delta Y-\beta\varepsilon)^2}{2\varepsilon}},
\end{equation}
expanding in Taylor series the left-hand-side of Eq.~(\ref{chapkol}) in powers of $\varepsilon$ and the right-hand-side in powers of $\Delta Y$ we
obtain

\begin{equation}\label{limitconti}
\begin{split}
&\Pi_{\varepsilon}+\varepsilon\frac{\partial \Pi_{\varepsilon}}{\partial S_n}+\frac{\varepsilon^2}{2}\frac{\partial^2\Pi_{\varepsilon}}{\partial S_n^2}+...=\\
&=\frac{1}{\sqrt{\pi}}\int_{-\infty}^{\frac{Y-\varepsilon\beta-Y_c}{\sqrt{2\varepsilon/a}}} dx \biggl(\frac{2\varepsilon}{a}\biggr)^\frac{n}{2} \left(x+\frac{\varepsilon\beta}{\sqrt{2\varepsilon/a}}\right)^n e^{-x^2}.
\end{split}
\end{equation}

It is worth noticing that  the Markovian probability distribution starting 
at the barrier value $Y_0=Y_C$ vanishes in $\sqrt{\varepsilon}$ while for $Y=Y_0=Y_c$ the first finite corrections to the probability distribution function  is in $\varepsilon$. At order one in $\varepsilon$ for $Y=Y_c$, this equation gives
\begin{equation}\label{epsilonfplanck}
\frac{\partial \Pi_{\varepsilon}}{\partial S_n}=\frac{a}{2}\frac{\partial^2 \Pi_\varepsilon}{\partial Y^2}-\frac{\beta}{2}\frac{\partial\Pi_\varepsilon}{\partial Y},
\end{equation}
which has the same form of than the Fokker-Planck equation associated with the Markovian solution for the DDB model in the continuous limit whose solution for a barrier at a generic point $Y_c$ is given by  Eq.~(3.11) in \cite{AchitouvCorasaniti2012}. In such case for $Y=Y_0=Y_c$ the drifting term stay in factor of Gaussian minus anti-Gaussian. Therefore we can assume that 

\begin{equation}\label{barrtobarr}
\Pi_{\varepsilon}(Y_c,Y_c,S_n)=C\varepsilon\frac{e^{-a\beta^2 S_n/2}}{S_n^{3/2}}.
\end{equation}

The value of the constant $C$ can be evaluating using the path-integral Eq.~(\ref{piepsilon}) for $n=2$ with $Y_0=Y_2=Y_c$ and
the explicit form of $W_0(..)$ given by Eq.~(A3) in \cite{CorasanitiAchitouv2011b}. Thus, equating Eq.~(\ref{barrtobarr})
on the left-hand-side to the result of the integral over $W_0(...)$ on the right-hand-side we obtain:
$C=\sqrt{\frac{a}{2\pi}}$.

\end{document}